\documentstyle[aas2pp4,epsf]{article}
\received{}
\accepted{}
\def\REF{\par\noindent\hangindent 20pt}
\def\msol{M$_\odot$\/}

\def\ltsima{$\; \buildrel < \over \sim \;$}
\def\simlt{\lower.5ex\hbox{\ltsima}}            % < over MMM
\def\gtsima{$\; \buildrel > \over \sim \;$}

\def\simgt{\lower.5ex\hbox{\gtsima}}            % > over MMM

\def\civ{{\sc{Civ}}$\lambda$1549\/}
\def\civnc{{\sc{Civ}}$\lambda$1549$_{\rm NC}$\/}

\def\cm3{cm$^{-3}$\/}
\def\hb{{\sc{H}}$\beta$\/}

\def\hbnc{{\sc{H}}$\beta_{\rm NC}$\/}

\def\ciii{{\sc{Ciii]}}$\lambda$1909\/}

\def\o4363{{\sc{[Oiii]}}$\lambda$4363\/}

\def\fe{{\sc{Fe}}\/}

\def\fe76087{{\sc [Fe vii]}$\lambda$6087\/}
\def\oiii{{\sc [Oiii]}$\lambda$5007}

\def\kms{km~s$^{-1}$}
\def\ergss{ergs s$^{-1}$\/}
\begin{document}
\title{\sc The Intermediate Line Region in AGN: a region ``pr\ae ter necessitatem''?}

\author{J. W. Sulentic\altaffilmark{1} and P. Marziani\altaffilmark{2}}

\altaffiltext{1}{Department of  Physics and Astronomy, University of
Alabama, Tuscaloosa, AL 35487; giacomo@merlot.astr.ua.edu}

\altaffiltext{2}{Osservatorio Astronomico di Padova, vicolo 
dell'Osservatorio 5, I-35122 Padova, Italy; marziani@pd.astro.it}

\keywords{Galaxies:  Seyfert -- Galaxies: Emission Lines --
Galaxies: Quasars -- Line:  Formation -- Line:  Profiles}

\begin{abstract}

As a consequence of improved S/N, spectral resolution and wavelength
coverage various authors have introduced, without strong justification,
new emitting regions to account for various emission line profile
differences in AGN.  The so-called CIV$\lambda$1549  intermediate line
region (ILR) appears to be especially ill-defined. We present
observational evidence that suggests the ILR is statistically
indistinguishable from the classical narrow line region (NLR). We
present the results of theoretical models showing that a smooth density
gradient in the NLR can produce CIV and Balmer emission lines with
different widths.  The putative ILR component has often been included
with the broad line profile in studies of CIV BLR properties. Failure
to account for the composite nature of CIV emission, and for the
presence of sometimes appreciable NLR CIV emission, has important
consequences for our understanding of the BLR.
\end{abstract}

\section{Introduction}

Observed profiles of different emission lines often show intriguing
differences in width and relative velocity. In some objects, there is a
rather strong correlation between line width and critical density or
ionization potential (De Robertis \&\ Osterbrock 1986).  These trends
have led to the concept of a  density gradient across the line emitting
regions, with the higher critical density (or higher ionization) lines
being emitted closer to the continuum source. The underlying assumption
is that the  velocity dispersion decreases with increasing distance
from the source of ionizing photons,  as expected for gas whose motion
is predominantly rotational or virialized.  Variations in line width
complicate attempts to separate the broad (BLR) and narrow (NLR) line
contributions in a source. The frequent observation  of a prominent
component with FWHM$\sim$ 1600 km s$^{-1}$ in the \civ\ (hereafter CIV)
profile complicates the interpretation of high ionization line (HIL)
properties in particular. In an attempt to take into account the
diverse widths of emission lines, some workers have proposed  the
existence of an ''Intermediate Line Region'' (ILR) (see e.g. Wills et
al. 1993). This emission component has been routinely treated as part
of the BLR emission in line profile studies either: a) because it is
believed to be part of the non-NLR gas or b) because a clear inflection
between the emitting components is not often seen in the CIV profile
making it difficult to subtract (see e.g.  Baldwin 1997, p. 92).

In Section 2 we discuss three observational results: 1) an unambiguous
CIV NLR  component with FWHM similar to [OIII]$\lambda$5007 is seen in
many Seyfert 2 galaxies, 2) a similar component is also identified in
some broad line AGN (with and without an inflection) although the
component is often broader with mean FWHM$\sim$ 900 km s$^{-1}$ broader
than FWHM H$\beta$ ($\approx$ FWHM [OIII]$\lambda$5007) and 3) except
for the broader profile width the  ``narrower'' CIV component shows
properties very similar to those ascribed to the classical NLR. Section
3 presents model calculations to show that the diversity of narrow line
profile widths can be  ascribed to a smooth gradient of physical
properties in the NLR, and that a larger width, $\sim$ 1000--2000 \kms,
is expected for CIV without invoking a new (ILR) emitting region.
Finally, in Section 4 we argue: a) that introduction of a new emitting
region is unjustified and b) that BLR properties inferred from the CIV
line are strongly dependent on whether or not a narrow component is
subtracted.

\section{An Unambiguous CIV NLR is Observed in Some AGN \label{detect}}

Most of the UV spectroscopic data for Seyfert 2 sources comes from the
IUE archive.  While the resolution of this data is limited, it shows
that the majority of Seyfert 2 galaxies exhibit a narrow line component
of CIV.  We assume, following unification ideas, that BLR CIV and
\hb\ are obscured in Seyfert 2's and that we see only NLR emission, the
same NLR in Seyfert 1 and 2 sources. Study of averaged spectra for a
sample of twenty mostly ``pure'' Seyfert 2 sources (Heckmann et al.
1995) show a significant narrow CIV line.   Detailed study of IUE
spectra for NGC 7674 and I Zw 92 (Kraemer et al. 1994) show an NLR line
intensity ratio CIV/\hb $\approx$ 2.0. NGC 1068 and 5643 are the only
Seyfert 2 sources in the HST archive with high s/n FOS spectra of the
CIV region.  NGC 1068 is the brightest example of a Seyfert 2 with a
confirmed hidden BLR (Antonucci \& Miller 1985).  Recent study of the
FOS spectra for NGC 1068 (Kraemer, Ruiz \&\ Crenshaw  1998) show that
it contains a strong \civ\  NLR component (partially resolved into
$\lambda\lambda$1548/1551 components) with FWHM$\approx$ 1100 $\pm$ 200
\kms\ . The narrow component of H$\beta$, and \oiii\, show profile FWHM
$\approx$ 1000 $\pm$ 200 \kms (Caganoff et al.  1991).  Clear evidence
for NLR CIV (FWHM=1180 \kms) is also seen in a bright Seyfert 1.5
galaxy, NGC 5548 (Goad \&\ Koratkar 1998). The NLR component in NGC
5548 remained constant while the BLR component went through large
changes, reenforcing the idea that independent NLR and BLR components
exist in that source.

The above cited observations make it clear that NLR CIV exists in many
lower luminosity AGN.  It is unambiguous in Seyfert 2 sources but can
be very confusing in sources that show BLR emission because a clear
NLR/BLR profile inflection is only occasionally observed. It has been
argued (Wills et al. 1993) that higher luminosity sources do not show
an NLR component.  This argument arose from profile analysis of a small
sample of high luminosity radio loud sources (Wills et al.  1993).  The
NLR was concluded to be weak or absent in these sources because no
significant profile component that matched the width of
[OIII]$\lambda$5007 was found. This interpretation was widely accepted,
in part, because reddening was assumed to be high in the NLR
(E(B-V)$\sim$0.5).  The above cited data suggests: 1) that the
reddening is frequently not so high (Kraemer, Ruiz \& Crenshaw 1998
derive a nuclear reddening of only E(B-V) =0.02 for NGC1068) and/or 2)
that CIV is amplified by some physical process (see section 3). The
current interpretation of narrow line spectra in NGC 1068 and 5548
points toward relatively high density and ionization parameter at the
inner edge of the NLR. A value of $\Gamma \sim 10^{-1.5}$  accounts for
the strength of several high ionization lines (Kraemer et al.  1998).
The dereddened NLR intensity ratio \civ/H$\beta$ is $\approx$ 14 for NGC
5548, and an inner NLR component with \civ/H$\beta$$\approx$ 30 appears
to be necessary.  Recently, Ferguson et al. (1997) interpreted the
strength of the strongest coronal lines observed in the spectra of
Seyfert galaxies in terms of emission from photoionized gas in a rather
dense region which could also be associated with the innermost NLR.

Our own line profile survey (Marziani et al.  1996: MS) of a
heterogeneous sample of 52 AGN found examples of both high and low
luminosity sources with a narrow CIV component comparable in width to
[OIII]$\lambda$5007.  However a considerable number of sources showed a
narrow component of CIV (and [OIII]$\lambda$4363, when it could be
resolved from H$\gamma$) that was broader than [OIII]$\lambda$5007 by
as much as 1500 km s$^{-1}$.  Figure 1 illustrates the distributions of
broad and narrow line widths found for CIV and H$\beta$ in the MS
survey.  The simplest interpretation is that both CIV NLR and BLR are
broader than the corresponding regions in H$\beta$. Additional support
for our NLR interpretation of the narrower CIV component comes from
observations of sources where the NLR is resolved. In the case of the
Seyfert 2 galaxy NGC 5643, the CIV NLR feature is obviously broader,
even in the extended emission line region (EELR: FWHM=1800$\pm$200
\kms), than either NLR HeII$\lambda$1640 or NLR \ciii.  PG2308+098 is
an example of a higher luminosity (z=0.43) source from the MS survey
that shows a clear inflection between the broad and, one of the
broadest observed, narrow CIV components. We derived a FWHM $\approx$
2630 \kms\ for the narrow CIV component considerably broader than FWHM
$\approx$ 700$\pm$100 \kms\ measured for \hb\ and \oiii\ in that
source.  Figure 2 shows a comparison between the CIV feature in the NGC
5643 EELR and the narrow CIV component derived for PG 2308+098. Figure
2 clearly shows that the width of the two narrow features are equal
within the observational errors. Narrow component widths as broad as
the one observed in PG2308+098 are clearly not unprecedented in
unambiguous NLR emitting regions.

\section{NLR Models: A Broader and Stronger \civ\ Line} 

We calculated photoionization models of the NLR and computed the
emergent spectrum using {\tt CLOUDY} (Ferland 1996). Our goal was to
account for FWHM differences as large as $\rm \Delta v_r \simlt 2000$
\kms (mean difference $\Delta$V= 900 km s $^{-1}$) between \hb\ and
CIV. We did not attempt sophisticated model innovations nor did we try
to fit the line profiles of a particular source. We attempted instead
to illustrate the implications of  recent robust results involving  the
input to photoionization calculations for the \civ\ and \hb\ line
profiles when the NLR line emission is not spatially resolved
(presently the case for the vast majority of AGN).  At z $\approx$ 0.2,
a 0.''25 aperture (typically employed for FOS observations) covers 1
kpc of spatial extent: thus most of the NLR should be included in a
measurement.

Input parameters for our models are listed in Table 1. We assumed
spherical symmetry with a radial density gradient  described by a
power-law, $\rm n_e = n_0 r_{pc}^{-s}$ \cm3\ from 1-10$^3$ pc. The
density at the outer edge of the NLR was set equal to $  10^3$ \cm3\ in
all cases. We assumed a  power-law dependence for the covering factor,
i.  e. $\rm f_{c,r} \propto r^{-m}$, with total NLR covering factor
0.1.  The first three models listed in Table 1 follow Netzer (1990)
with a column density $\rm N_c  \propto r^{-2/3s}$. The last four
models assumed that $\rm N_c \approx 10^{23}$ cm$^{-2}$ = constant
throughout the NLR (Column 3 lists adopted values of m). The inclusion
of the effects of appreciable mechanical heating appear to be necessary
for realistic modeling of the NLR. HST narrow-band images and radio
maps have show a close relationship between radio ejecta and NLR
emitting plasma in several nearby Seyfert galaxies (Capetti et al.
1999; Falcke et al. 1998). A crude estimate of the heating rate
$\epsilon$\ can be obtained by assuming that a fraction $\rm f_P$\
$\approx 0.1$ (approximately equal to the covering factor $\rm f_c$) of
the jet energy flux $\Pi$\ is converted into heating of the NLR gas.
Assuming an exponential radial dependence of the heating rate we can
write $\rm \epsilon = \epsilon_0 f_{c,0.1} R_{out,1 kpc}^{-3} \Pi_{45}
\exp(-(r-r_0)/r_s) $\ergss\ \cm3, where we have parameterized the
dependence on outer radius (R$\rm _{out}$), $\rm f_c$, and $\Pi$
(Column 4 lists the scale factor).  The values $\rm r_S = 1$ and 10pc
correspond, respectively, to a heating rate per unit volume at 1 pc
$\log \epsilon_0 \approx  -9.38$ (which is neglible) and $\approx
-6.37$ (which provides significant enhancement of electron temperature,
in a cocoon within 2 pc). The latter cases is also more realistic
because recent hydrodynamic simulations show that most of the line
luminosity is concentrated in dense clouds surrounding the jet (Steffen
et al. 1997).  However, recent observations (Villar-Martin et al.
1999) of high redshift radio-loud and quiet sources reveal FWHM$\simgt$
1000 \kms emission lines in extended gas both near and far from radio
jets.  The jet energy flux is a poorly known quantity, and its actual
relevance depends on the details of the jet/cloud interaction.
Significant heating of clouds can however be expected for a wide range
of AGN including radio quiet objects like Seyferts (in agreement with
the analysis of Bicknell et al. 1998 for Seyfert 2 galaxies).

The gas is assumed to move with velocity $\rm v(r) = (2100
r^{-1/2} + 0.1 r)$\ i.e., at the  local virial velocity plus a
component due to galactic rotation.  A velocity dispersion $\rm \Delta
v \approx 4150 r^{-0.5}_{1~pc} M^{0.5}_9$ \kms\ is expected at a
distance r equal to 1 pc (which we assume as the inner edge of the NLR;
1 pc $\approx$ 10$^4$ gravitational radii for a $10^9$ \msol\ black
hole), if the gas motion is rotational or virial.  We focus our profile
predictions on four NLR lines: CIV, \hb, \oiii, and \fe76087. The
widths reported in Table 1 (Cols.  5--8) are the weighted average of
velocity over line emissivity per unit area $\rm \zeta_r$ (i.e., $\rm <
v > = {\int v_r f_{c,r} \zeta_r dV}/{\int f_{c,r} \zeta_r dV}$).
Observed NLR linewidth for \oiii, H$\beta$, and \fe76087\ are
reproduced by the models (FWHM(\oiii) $\approx$ FWHM(H$\beta$) $<$
FWHM(\fe76087) $\sim$ 1000 \kms), with models 5 and 7 a slightly larger
with of \oiii\ with respect to \hb, as it is often observed.  The most
relevant result  is that $< v >$\  for \civ\ appears to be significantly
larger than that of any other line considered, with $< v > \sim
1000-2500$\kms$ \simgt$ 1000 \kms\ in all cases.  \civ\ emission is
favored by (a) the larger ionization parameter in the innermost NLR,
and (b) mechanical heating.

A more realistic case might involve  cylindrical symmetry, for example,
a cylindrical NLR volume with radius 1 kpc and total height 3.4 kpc
(the values appropriate for Mark 50; Falcke et al.  1998), where the
axis of symmetry corresponds to the radio axis.  No substantial
difference from the spherically symmetric models is expected as far as
$< v >$\ is concerned,  if similar assumptions about gas motions are
made. However, the cylindrical symmetry case may give: (a) a larger
\civ\ flux with respect to the spherically symmetric case and (b) a
possibly broader \civ\ due to the pressure exerted by the jet, which
could accelerate dense gas clumps.  Mechanical heating may also provide
a suitable mechanism for dust destruction.  More sophisticated
modelling (e. g.  Steffen et al. 1997; Bicknell et al. 1998) is beyond
the scope of this paper.

\section{Results}
\subsection{The Intermediate Line Region: is it pr\ae ter necessitatem?
\label{ilr}}

Introduction of an additional (e.g. ILR) emitting region can be
justified by the following (not mutually exclusive) circumstances:  1)
spatial segregation (e.g. region associated with a jet) and/or 2) an
abrupt discontinuity in physical properties (e.g. optical thickness or
cloud instability at some distance). This is the physical statement of
Occam's razor ({\it multiplicanda non sunt pr\ae ter necessitatem}).
The above circumstances describe the justification for the concept of
distinct BLR and  NLR regions where suppression of collisional lines
yield  different spectra. However ``ILR'' emission and coronal lines
($<v>\approx 1-2\times10^3$ \kms) can be produced in the inner part
of the NLR which, even if associated with an inner sheet adjacent to
radio plasma, does not require introduction of a new distinct and
discontinuous emitting region.

Part of the confusion over the identification of different line
emitting components may have arisen from a difference in nomenclature.
Wills et al. (1995) and Brotherton et al. (1994a,b) refer to the
broader and narrower  CIV  components  as very broad line (VBLR) and
intermediate line (ILR) respectively.  However they also inferred
(Brotherton et al. 1994b) that the ILR properties (reverberation
distance from continuum source, density, covering factor) are all
similar to the standard NLR values. The ILR may be an ``inner part of
the NLR'' but it is NLR and not BLR gas.  Our models strengthen this
view. The VBLR designation is also confusing because this name has
already been applied to an even broader component sometimes seen in the
HeII$\lambda$4686 feature (Ferland et al. 1990; Marziani \&\ Sulentic
1993).

\subsection{BLR Implications}

The model calculations support our two-component NLR-BLR interpretation
of the CIV and \hb\ profiles. Of course greater broad line profile
complexity is sometimes observed, with double-peaked Balmer line
profiles an extreme example. The relative strength and frequency of
occurence of multiple BLR components is not yet well established.
However one interprets the narrower CIV feature, there is no
justification for treating it as BLR gas. A number of recent studies
involving the CIV BLR have analysed profile data without the benefit of
subtracting any NLR component (e.g.  Wills et al. 1993; Brotherton et
al. 1994ab; Corbin \& Francis 1994; Corbin \& Boroson 1996).

Does it matter if an NLR component is subtracted from CIV?  Our results
suggest that the answer is a strong ``yes''. The NLR component
subtracted from CIV (in MS) represented 2-20\% of the total flux in
most cases.  This flux was concentrated in the narrow peak of the line
and therefore affected disproportionately determinations of the half
maximum level of the broad line profile. The result of nonsubtraction
yields line parameters surprisingly different from those obtained by
MS.  Figure 3 shows a comparison of CIV centroid line shifts (relative
to [OIII]5007) for sources in common between MS and Corbin \& Boroson
(1996). A similar comparison also reveals that the CIV BLR profile
width (FWHM) is measured, on average, 40\%\ smaller when an NLR
component is not subtracted.  High ionization line BLR properties
inferred from CIV, arguably the least confused HIL line, are dependent
on whether or not an NLR is subtracted.

If an NLR component is subtracted, MS survey results suggest that:  1)
radio-loud (RL) AGN show fundamentally different BLR
geometry/kinematics from the radio quiet (RQ) majority, 2) in RQ--BLR
CIV is blueshifted (up to 3000 km s$^{-1}$) with respect to the local
rest frame while H$\beta$ show small red and blue velocity shifts (up
to 600 km s$^{-1}$) about that frame and 3) In RL--BLR CIV shows zero,
or small red and blue (up to 1000 km s$^{-1}$), velocity excursions
relative to the rest frame while H$\beta$ shows larger predominantly
red (but sometimes blue) shifts (up to 3000 km s$^{-1}$) relative to
that frame. These effects must be tested with samples that cover a
larger source luminosity range.

\acknowledgements

PM acknowledges financial support by Italian Ministry for University and
Research (MURST) under grand  Cofin98-02-32.

\newpage
\section{References}
\REF
Antonucci, R. and Miller, J. 1985, ApJ, 297, 621
\REF
Baldwin, J. 1998, In proceedings of  Emission Lines in Active
Galaxies, eds. B. Peterson et al., ASP Conf. Ser.,  113, 80
\REF
Bicknell, G. V., Dopita, M. A., Tsvetanov, Z. I., \&\ 
Sutherland, R. S. 1998, ApJ, 495, 680
\REF
Brotherton, M. S., Wills, B. J., Steidel, Ch., \&\ Sargent W. L. W., 1994a,  ApJ,  423, 131
\REF
Brotherton, M. S., Wills, B. J., Francis, P. J., \&\  Steidel, Ch. C. 1994b,  ApJ,  430, 495
\REF
Caganoff, S., et al. 1991, ApJ 377, 9 
\REF
Capetti, A., et al. 1999,  MemSAIt, in press
\REF
Corbin, M. \&\ Francis, P. 1994,  AJ,  108, 2016
\REF
Corbin M., R., \&\ Boroson, T. 1996,  ApJ Suppl,  107, 69
\REF
De Robertis, M. M., \&\ Osterbrock, D. E. 1986, ApJ, 301, 727
\REF
Falcke, H., Wilson, A. and Simpson, C. 1998, ApJ, 502, 199
\REF
Ferguson, J., Korista, K. and Ferland, G. 1997, ApJ Suppl, 110, 287
\REF 
Ferland, G., Korista, K. and Peterson, B. 1990, ApJ, 363, L21
\REF
Ferland, G. J. 1996, Hazy: a Brief Introduction to CLOUDY, Univ. Kentucky, 
Dept. Phys. \&\ Astron., internal report 
\REF
Goad, M., \&\ Koratkar, A. 1998, ApJ 495, 718
\REF
Heckmann, T. et al. 1995,  ApJ,  452, 549
\REF
Kraemer, S., et al. 1994,  ApJ,  435, 171
\REF
Kraemer, S., Crenshaw, D., Filippenko, A. and Peterson, B. 1998, ApJ, 499, 719
\REF
Kraemer, S., Ruiz, J. and Crenshaw, M. 1998, ApJ, 508, 232
\REF
Marziani, P. \&\ Sulentic, J. W. 1993,  ApJ,  409, 612
\REF
Marziani, P., Sulentic, J. W., Dultzin-Hacyan, D., Calvani, M. \&\
Moles, M. 1996,  ApJ Suppl,  104, 37 (MS)
\REF
Netzer, H., 1990, in Active Galactic Nuclei, Saas Fee Advanced Course 20, R.D.
Blandford, H. Netzer, L. Woltjer (Berlin:Springer), p. 57
\REF
Steffen, W., Gomez, J. L., Raga, A. C., \&\ Williams, R. J. R. 1997, ApJ 491, L73
\REF
Villar-Martin, M., Binette, L. and Fosbury, R. 1999, A\&A, in press
\REF
Wills, B. et al. 1993,  ApJ,  415, 563
\REF
Wills, B. J., Thompson, K. L.,  Han, M., Netzer, H., Wills, D., Baldwin, J. A.,
Ferland, G. J.,  Browne, I. W. A., \&\  Brotherton, M. S. 1995, ApJ 447, 139

\newpage
\begin{deluxetable}{cccccccc}
\tablewidth{42pc}
\tablecaption{Model Computations \label{contacci}}
\tablehead{
\colhead{Model}  & \colhead{$\rm \log(n_0)$}
 &
\colhead{m} & \colhead{$\rm r_s$\tablenotemark{1}}
& \colhead{$< v >$}
& \colhead{$< v > $} & \colhead{$< v>$} & \colhead{$< v
> $ } \\
&                  &       &   &\colhead{\hbnc} & \colhead{\oiii} &
\colhead{\fe76087} &
\colhead{\civnc\tablenotemark{2}}\\
& \colhead{(\cm3)} &  & (pc) & \colhead{(\kms)} & \colhead{(\kms)} &
\colhead{(\kms)} & \colhead{(\kms)}\\
\colhead{(1)} & \colhead{(2)} &
\colhead{(3)} & \colhead{(4)} & \colhead{(5)} & \colhead{(6)} &
\colhead{(7)} & \colhead{(8)} }
\startdata
1 &  8   & $\frac{7}{18}$ & 10 &  490 & 450 & 1030 & 1370 \\
2 &  7  & $\frac{11}{18}$ & 10 & 420 & 310 & 950 & 1250\\
3 &  8   & $\frac{7}{18}$ & 1 &   720 & 440 & 1160 &  2680  \\
4 &  8   & $\frac{3}{2}$ & 10 &  370 & 340 & 980 & 1380 \\
5 &  7   & $\frac{3}{2}$ & 10 & 370 & 490 & 910 & 1210 \\
6 &  8   & $\frac{3}{2}$ & 1 & 380 & 330& 1020 & 1450  \\
7 &  7   & $\frac{3}{2}$ & 1 & 360 & 520 & 925 & 1950 \\
%Spherical &  7  & $\frac{4}{3}$ & $\frac{11}{18}$ & 1 & \\
\enddata
\tablenotetext{1}{Scale factor for mechanical heating, assumed to
decrease exponentially from $\rm r_0$ ~~(see text).}
\tablenotetext{2}{Width of single line.}
\end{deluxetable}

\newpage

\begin{figure} 
%\figurenum{1}
\plotone{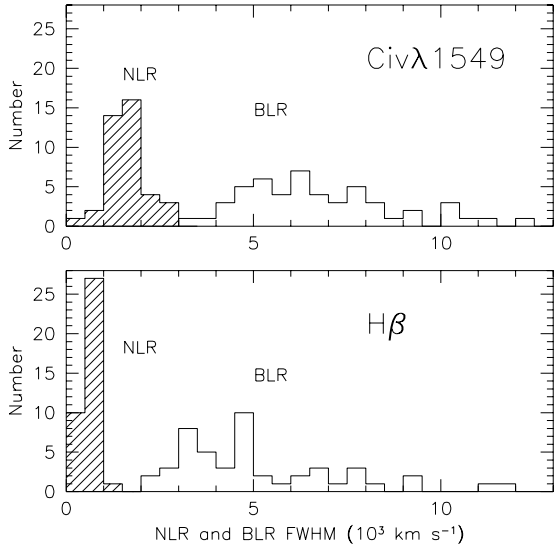}
\caption[1]{FWHM distributions for the broad and narrow components of CIV 
and H$\beta$. Data from the MS survey (Marziani et al. 1996).}
\end{figure}

\begin{figure} 
%\figurenum{1}
\plotone{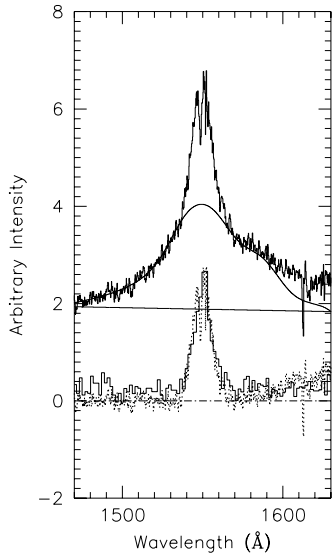}
\caption[1]{CIV$\lambda$1549 profiles for: i) PG 2308+09 (upper) -
smooth line shows fits to the BLR and continuum components and ii) NGC
5643 EELR (lower) - scaled and without continuum subtraction. NLR
component of PG2308+09 superposed (dotted). Note the similar profile 
widths.}
\end{figure}

\begin{figure} 
%\figurenum{1} 
\plotone{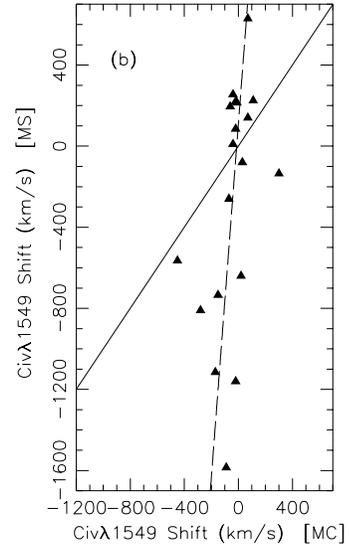} 
\caption[1]{CIV BLR profile shift measurements (at FWHM level) for
sources in common between the MS survey and Corbin \&\ Boroson (1996)
(MC). Dashed line is an  unweighted least squares fit and the thin line
the locus of equal shifts. MC measures are consistent with no shift
within the errors ($\pm 200$ \kms).} 
\end{figure}

\end{document}